\documentstyle[aps,preprint]{revtex}
\input epsf

\begin{document}

\title{Determining the structure of the superconducting gap \\
in Cu$_2$O$_3$ 2--leg ladder materials \\}

\author{ N.\ Bulut and D.J.\ Scalapino }

\address{
Department of Physics, University of California \\
Santa Barbara, CA 93106--9530} 

\maketitle 
\begin{abstract}
Superconductivity has been recently observed in 
Sr$_{0.4}$Ca$_{13.6}$Cu$_{24}$O$_{41.84}$
which contains quasi--one--dimensional Cu$_2$O$_3$
2--leg ladders.
If, as suggested by some theories, the superconductivity 
arises from these 2--leg ladders,
it will be important to determine the structure of the 
superconducting gap.  
In particular, does the gap on a 2--leg ladder change 
sign when one goes from the bonding to antibonding fermi
surface points?
Here we carry out phenomenological calculations of nuclear
relaxation rates and inelastic neutron scattering intensity
in order to provide estimates of the experimental resolution that
will be required to determine the structure of the
superconducting gap associated with an array of 
weakly coupled 2--leg ladders.
\end{abstract}

\pacs{PACS Numbers: }

Recently Uehara {\it et al.} \cite{Uehara} have reported 
the observation of superconductivity at 12K in
Sr$_{0.4}$Ca$_{13.6}$Cu$_{24}$O$_{41.84}$ under 3GPa of 
pressure.
This material consists of alternate planes containing Cu$_2$O$_3$
2--leg ladders and single CuO$_2$ chains separated by Sr--Ca layers.
The appearance of superconductivity is consistent with theoretical
predictions of pairing in doped 2--leg ladders 
\cite{Dag,Rice,Gopalan,Noack}.
However, further measurements to determine whether pairing in fact 
occurs on the 2--leg ladders and the nature of this pairing will 
determine to what extent the observed superconductivity is 
related to the theoretical predictions.
In particular, according to the theoretical calculations, 
the pairing on the 2--leg ladders is $d_{x^2-y^2}$--like in that 
the sign of the gap at the two bonding band fermi surface points 
is opposite to that at the antibonding band fermi surface points.
Alternatively, in a stronger coupling local view, this means 
that in the real space superposition of the singlets that makes up the pair,
there is a minus sign difference between a singlet across a rung and a 
near neighbor singlet along a leg.
Now, as opposed to the two--dimensional case in which 
the node associated with a gap having $d_{x^2-y^2}$--like symmetry
gives rise to a linear low temperature dependence of the 
Knight shift and the penetration depth, the 2--leg ladder is nodeless
and, for example, the low temperature Knight shift will 
decay exponentially on a scale set by the magnitude of the gap.
Knight shift measurements will thus be unable to distinguish 
a gap which changes sign on the bonding and antibonding 
fermi surfaces from one which does not.
However, nuclear relaxation times as well as the inelastic 
neutron scattering intensity depend upon coherence factors and thus 
should in principle allow one to determine the relative sign 
difference of the gap.
Here we present the results of a phenomenological calculation 
of these quantities in order to estimate what experimental 
resolution will be required to determine the gap structure on the 
2--leg ladder.
We will compare the results for the 
$d_{x^2-y^2}$--like gap with those of an $s$--like gap which has 
the same sign at the 
bonding and antibonding fermi surface points \cite{Like}.

Both the nuclear relaxation times and the neutron scattering intensity 
depend upon the magnetic susceptibility.
Here we use a phenomenological RPA--BCS form for 
$\chi({\bf q},\omega)$,
which we previously introduced to describe 
the layered cuprates \cite{BS}.
In this approach, the short range antiferromagnetic correlations are
taken into account by using the RPA form 
\begin{equation}
\chi({\bf q},\omega) = 
{\chi_0({\bf q},\omega) \over 
1-\overline{U} \chi_0({\bf q},\omega) }
\label{eq:chi}
\end{equation}
with $\overline{U}$ as a parameter.
The superconducting correlations are modeled by using a BCS form for
$\chi_0({\bf q},\omega)$
\begin{eqnarray}
\chi_0({\bf q},\omega)
={{1\over N}} \sum_{{\bf p}}
& &\Biggl\{
{{1\over 2}} 
\left[1+ 
{ \varepsilon_{{\bf p}}\varepsilon_{{\bf p}+{\bf q}} +
\Delta_{{\bf p}} \Delta_{{\bf p}+{\bf q}} 
\over 
E_{{\bf p}} E_{{\bf p}+{\bf q}} }
\right]
{f(E_{{\bf p}+{\bf q}})-f(E_{\bf p}) \over 
\omega-(E_{{\bf p}+{\bf q}}-E_{\bf p})+i\delta} \nonumber \\
& &+{{1\over 4}} 
\left[1-
{ \varepsilon_{{\bf p}}\varepsilon_{{\bf p}+{\bf q}} +
\Delta_{{\bf p}} \Delta_{{\bf p}+{\bf q}} 
\over 
E_{{\bf p}} E_{{\bf p}+{\bf q}} }
\right]
{1-f(E_{{\bf p}+{\bf q}})-f(E_{\bf p}) 
\over 
\omega +(E_{{\bf p}+{\bf q}}+E_{\bf p})+i\delta} \\
\label{eq:chi0}
& &+{{1\over 4}} 
\left[1-
{ \varepsilon_{{\bf p}}\varepsilon_{{\bf p}+{\bf q}} +
\Delta_{{\bf p}} \Delta_{{\bf p}+{\bf q}} 
\over 
E_{{\bf p}} E_{{\bf p}+{\bf q}} }
\right]
{f(E_{{\bf p}+{\bf q}})+f(E_{\bf p})-1
\over 
\omega-(E_{{\bf p}+{\bf q}}+E_{\bf p})+i\delta} \nonumber
\Biggr\}.
\end{eqnarray}
\narrowtext
Here $\Delta_{{\bf p}}$ is the superconducting gap,
$\varepsilon_{{\bf p}}$ is the bare quasiparticle dispersion,
$E_{\bf p}=\sqrt{ \varepsilon_{\bf p}^2 + \Delta_{\bf p}^2 }$,
and $f$ is the usual fermi factor.
The quasiparticle dispersion of the two--chain tight binding 
model with the hopping matrix element $t$ along the chains
and $t_{\perp}$ along the rungs is
\begin{equation}
\varepsilon_{{\bf p}}=-2t\cos{p_x}-t_{\perp}\cos{p_y}-\mu,
\label{eq:ep}
\end{equation}
where $\mu$ is the chemical potential.
For simplicity, we will consider the case of 
isotropic hopping $t_{\perp}=t$.
In order to model a $d_{x^2-y^2}$ superconductor 
we use the simple form
\begin{equation}
\Delta_{{\bf p}} = {\Delta_0\over 2} 
(\cos{p_x} - \cos{p_y}).
\label{eq:deltad}
\end{equation}
Since we are interested in the low frequency magnetic 
response of the system,
the only important feature of this gap form is that 
$\Delta_{{\bf p}}$ is finite and has different signs
on the bonding ($p_y=0$) and 
antibonding ($p_y=\pi$) fermi surface points.
One could have modeled the $d_{x^2-y^2}$ gap just as well 
by using a form which is $+\Delta_0$ on the bonding band
and $-\Delta_0$ on the antibonding band.
In order to distinguish the $d_{x^2-y^2}$ gap, 
we will also make comparisons with results obtained 
from an $s$ type of gap having the form
\begin{equation}
\Delta_{{\bf p}} = \bigg|{\Delta_0\over 2} 
(\cos{p_x} - \cos{p_y})\bigg|.
\label{eq:deltas}
\end{equation}
In obtaining the following results we have assumed a
mean field superconducting transition temperature $T_c=0.1t$,
electronic filling $\langle n\rangle=0.85$ and 
$\overline{U}=1.5t$.
We have also assumed a BCS temperature dependence for the magnitude 
of the gap with 
$2 {\rm max}\,(\Delta({\bf p}_{\rm F}))=7T_c$, where 
${\rm max}\,(\Delta({\bf p}_{\rm F}))$ is the maximum value of the
gap at the fermi points.
Our conclusions will not depend on small variations 
in these parameters.

We will study the longitudinal relaxation rate $T_1^{-1}$ for 
$^{63}$Cu and $^{17}$O nuclei.
The nuclear relaxation rate $T_1^{-1}$ is given by 
\begin{equation}
{1\over T_1} = {T\over N} \sum_{{\bf q}}
|A({\bf q})|^2 
\lim_{\omega\rightarrow 0}
{ {\rm Im}\,\chi({\bf q},\omega) \over \omega},
\label{eq:rate1}
\end{equation}
where $|A({\bf q})|^2$ is the form factor of the corresponding nuclei.
For a Cu$_2$O$_3$ ladder,
the $T_1^{-1}$ response of the oxygen nuclear spins
depends on whether they are located on a chain (along the $x$ axis)
or on a rung (along the $y$ axis).
We will assume that the hyperfine hamiltonian
determining the relaxation of the oxygen nuclear spins is dominated by 
a transferred hyperfine coupling to the electronic spins
centered at the two neighboring Cu sites.
The resulting hyperfine
form factor of a chain $^{17}$O nuclei is given by 
\begin{equation}
|A_{\rm O}^{\rm chain}({\bf q})|^2 = 4|A_{\rm O}|^2\cos^2(q_x/2),
\label{eq:aqchain}
\end{equation}
while that of a rung $^{17}$O is 
\begin{equation}
|A_{\rm O}^{\rm rung}({\bf q})|^2 = 4|A_{\rm O}|^2\cos^2(q_y/2).
\label{eq:aqrung}
\end{equation}
We note that for $q_y=\pi$, which corresponds to scatterings between
the bonding and antibonding bands, the form factor of the rung 
$^{17}$O nuclei vanishes.
Hence the relaxation rate of the rung $^{17}$O
does not probe the relative phase difference of 
the gap on the bonding and antibonding fermi points.
Thus for this nuclear spin the $d_{x^2-y^2}$ and $s$--wave gaps
will give the same result.
However, $T_1^{-1}$ for the chain $^{17}$O are influenced by
the relative phase of the superconducting gap.
For the $^{63}$Cu nuclei, we will assume only an onsite 
hyperfine coupling and use
\begin{equation}
|A_{\rm Cu}({\bf q})|^2 = |A_{\rm Cu}|^2.
\label{eq:aqcu}
\end{equation}
This is a reasonable approximation since 
in the Cu$_2$O$_3$ ladder compounds
the magnitude of the 
transferred hyperfine coupling to the near--neighbor $^{63}$Cu
sites is small compared to the 
onsite hyperfine coupling \cite{Ishida,SDS}.

\begin{figure}
\centerline{\epsfysize=7.2cm \epsffile[-30 184 544 598] {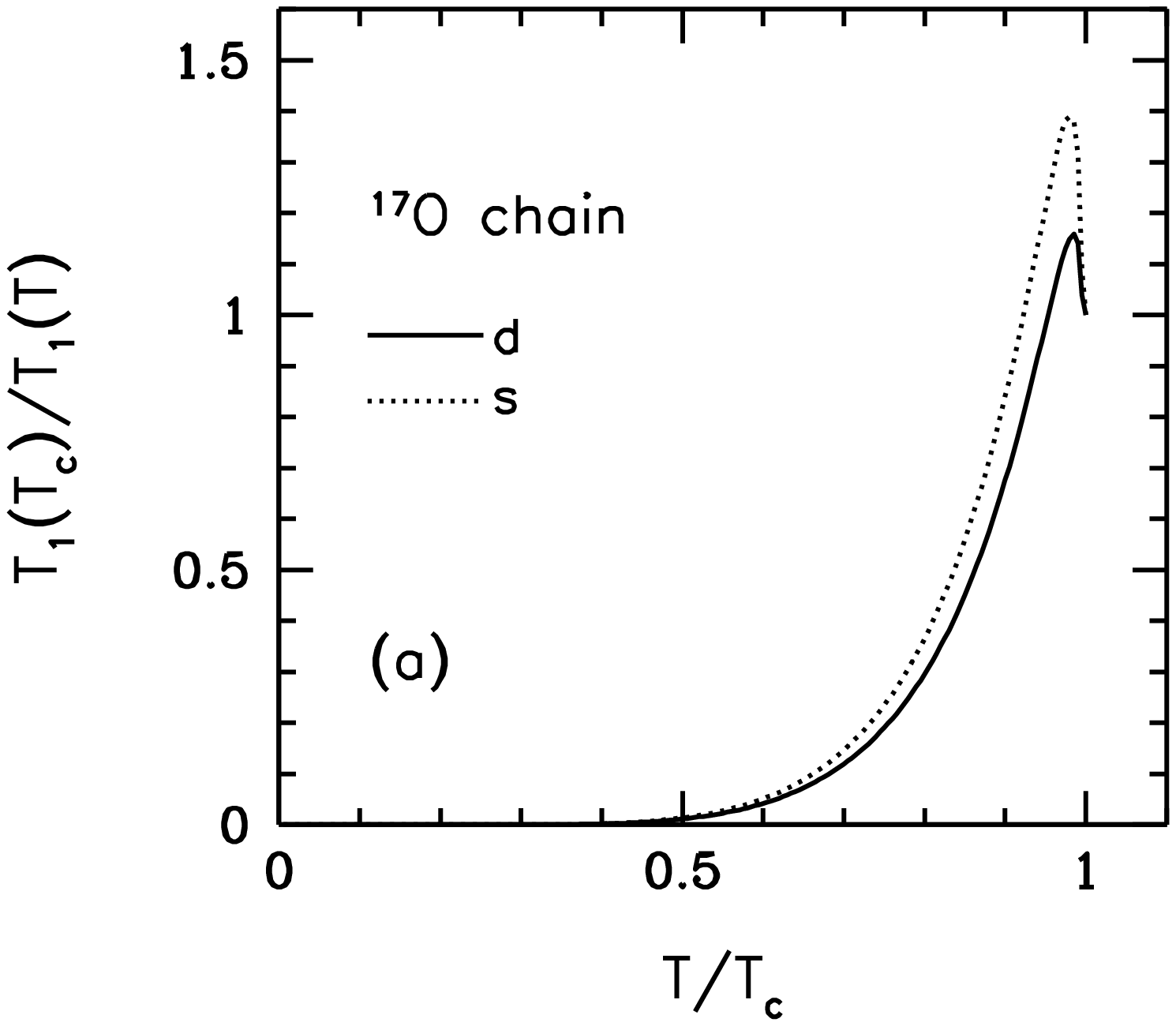}
\epsfysize=7.2cm \epsffile[98 184 672 598] {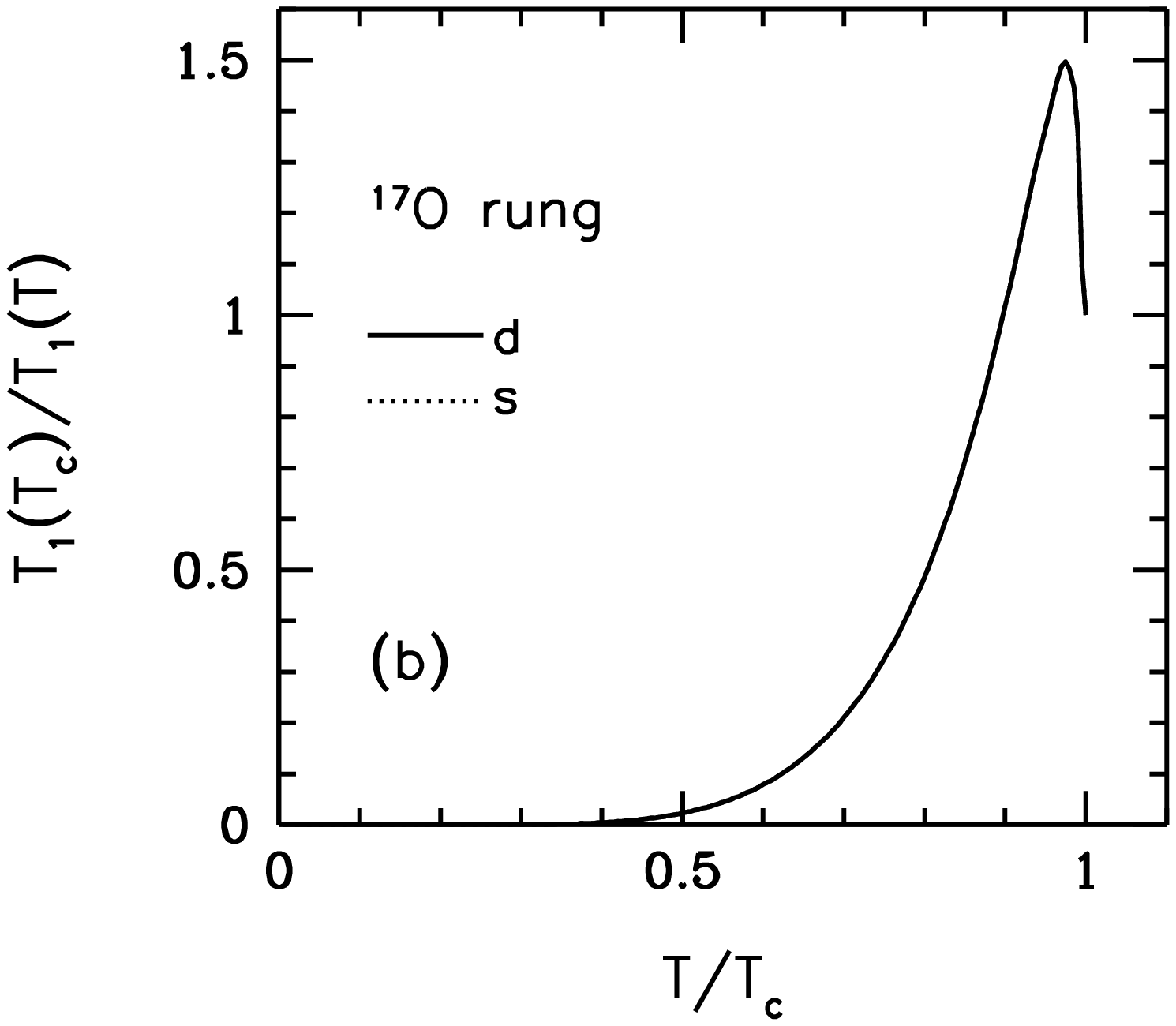}}
\centerline{\epsfysize=7.2cm \epsffile[18 184 592 598] {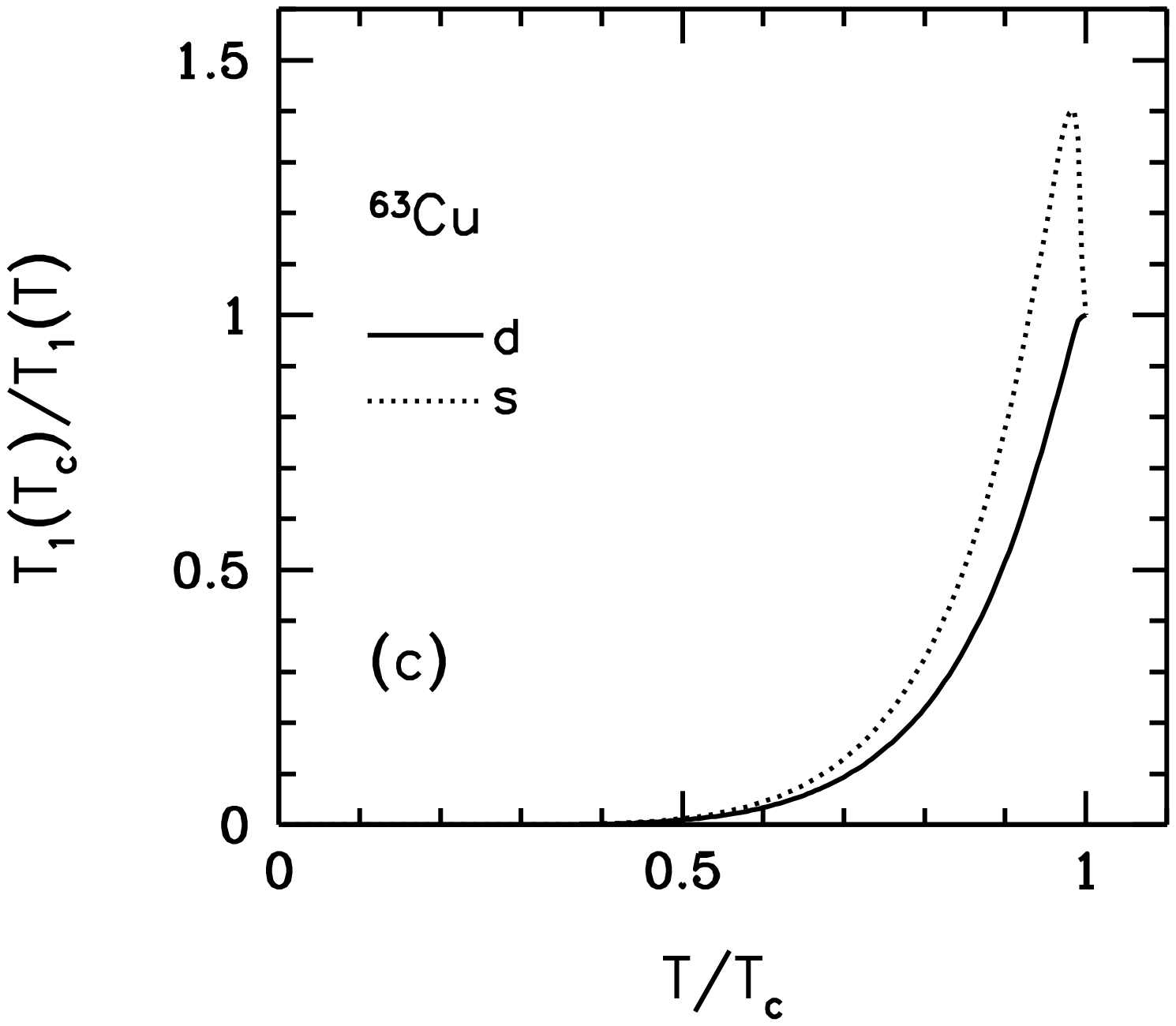}}
\vspace{1.4cm}
\caption{
Temperature dependence of the NMR rate $T_1^{-1}$ for 
(a) the chain $^{17}$O, 
(b) rung $^{17}$O and 
(c) $^{63}$Cu nuclei obtained using the $d_{x^2-y^2}$ and $s$
type of gaps.
For the rung $^{17}$O nuclei, these two gap structures
yield identical results.
The $s$--wave gap yields a Hebel--Slichter peak 
in all three of the $T_1^{-1}$ rates.
For the $d_{x^2-y^2}$ gap, the $T_1^{-1}$ of the $^{63}$Cu nuclei 
does not have a Hebel--Slichter peak, while the $T_1^{-1}$ 
of the chain $^{17}$O nuclei has a peak which is reduced with respect 
to that of the rung $^{17}$O nuclei.
\label{fig:rate1}}
\end{figure}

Figures \ref{fig:rate1}(a)--(c) show the temperature dependence of 
the nuclear relaxation rate $T_1^{-1}$ for the chain and 
rung $^{17}$O, and $^{63}$Cu nuclei.
For the $s$--wave gap, there is a Hebel--Slichter coherence peak 
right below $T_c$ for all three of the relaxation rates.
As it is well known, the Hebel--Slichter peak is due to the 
non--vanishing of the coherence factor of the first term 
in Eq.~(\ref{eq:chi0}) and the singularity in
the superconducting density of states.
For the $d_{x^2-y^2}$ gap, the $T_1^{-1}$ of the rung $^{17}$O is
identical to that of the $s$--wave gap.
This is because its form factor vanishes at $q_y=\pi$,
and, also, only the magnetic fluctuations with momentum transfer
$q_y=\pi$ are sensitive to the relative sign of the $d_{x^2-y^2}$
gap, while the $q_y=0$ fluctuations are not.
On the other hand, the form factor of the chain $^{17}$O nuclei 
allows comparable 
contributions to the $T_1^{-1}$ rate from both the $q_y=0$ and
$q_y=\pi$ magnetic fluctuations.
Consequently, for a $d_{x^2-y^2}$ gap, the $T_1^{-1}$ of the 
chain $^{17}$O has a reduced peak with respect to that of the 
rung $^{17}$O \cite{footnote1}.
At $T=T_c$, the $T_1^{-1}$ rate of the $^{63}$Cu nuclei is 
dominated by antiferromagnetic fluctuations with $q_y=\pi$,
and hence it does not have a Hebel--Slichter peak 
for a $d_{x^2-y^2}$ gap.
These results show that the $d_{x^2-y^2}$ and $s$--wave 
gaps yield quite different temperature dependences for the 
$^{17}$O and $^{63}$Cu $T_1^{-1}$ rates, hence measurements 
of these relaxation rates can be used to identify the structure
of the superconducting gap function.

Another probe of the spin dynamics in the superconducting state 
is the transverse nuclear relaxation rate $\tau^{-1}$ 
given by \cite{Pennington}
\begin{equation}
{1\over \tau^2} = 
{1\over N}\sum_{{\bf q}} |A({\bf q})|^4 \chi^2({\bf q},0) 
-
\bigg({1\over N}\sum_{{\bf q}} |A({\bf q})|^2\chi({\bf q},0)
\bigg)^2
\label{eq:rate2}
\end{equation}
for a nuclear spin which has a form factor $|A({\bf q})|^2$.
Figures 2(a)--(c) show the temperature dependence of $\tau^{-1}$ for 
the three nuclei in the superconducting state.
For the rung $^{17}$O, $\tau^{-1}$ 
is identical for the 
$d_{x^2-y^2}$ and $s$ type of gap structures.
It would be difficult to identify the structure of the 
gap from $\tau^{-1}$ of the chain $^{17}$O also,
since it is similar for both types of gaps.
However, the results for the $\tau^{-1}$ of the $^{63}$Cu nuclei
are quite distinct.
While there is approximately a 60\% 
decrease in the 
$s$--wave result for $\tau^{-1}$
as the temperature is lowered in the superconducting state,
for a $d_{x^2-y^2}$ gap it drops by only 15\%.
This reflects the fact that 
in the superconducting state the antiferromagnetic 
correlations are more strongly suppressed for an $s$--wave gap than 
for a $d_{x^2-y^2}$--wave gap \cite{footnote2}.
Hence measurements of $\tau^{-1}$ for the
$^{63}$Cu nuclei would be helpful in determining the structure 
of the superconducting gap.

Inelastic neutron scattering is a direct probe of the 
spin fluctuation spectral weight ${\rm Im}\,\chi({\bf q},\omega)$.
The wavevectors which connect the bonding and 
antibonding fermi surface points,
${\bf p}_{\rm Fb}=(\pm 0.6\pi,0)$ and 
${\bf p}_{\rm Fa}=(\pm 0.25\pi,\pi)$,
are particularly helpful in 
distinguishing between the $d_{x^2-y^2}$ and $s$ type of gap structures.
In Figure 3 we show results for one such wave vector,
${\bf q}^*=(0.85\pi,\pi)$.
Here we see that for the $d_{x^2-y^2}$ gap
there is a resonance at the gap edge
$\omega=|\Delta({\bf p}_{{\rm Fa}})| + |\Delta({\bf p}_{{\rm Fb}})|$.
This resonance is due to the non--vanishing of the coherence factor
for a $d_{x^2-y^2}$ gap which is 
associated with quasiparticle creation,
and to the RPA form of Eq.\,(\ref{eq:chi}).
On the other hand, 
for an $s$--wave gap this coherence factor vanishes,
and the quasiparticle creation process at the gap edge
is suppressed.

In summary, we have shown how one can obtain information on
the structure of the superconducting gap for a Cu$_2$O$_3$ ladder
system from NMR and magnetic neutron scattering experiments.
This problem is of interest because of the recent discovery of 
superconductivity in the ladder material
Sr$_{0.4}$Ca$_{13.6}$Cu$_{24}$O$_{41.84}$.
It is especially important to know whether the 
superconducting gap has different signs 
on the bonding and antibonding fermi surface points,
since this is relevant to our understanding of the nature of 
the pairing interaction.
Here, we have seen that measurements of the NMR rates $T_1^{-1}$ and
$\tau^{-1}$ and the inelastic neutron scattering intensity can provide
useful information on the structure of the superconducting gap
in this new compound.
For the $d_{x^2-y^2}$ gap, we expect a Hebel--Slichter
peak for the rung $^{17}$O nuclei and a reduced peak 
for the chain $^{17}$O, but not for the $^{63}$Cu nuclei.
An $s$ type of gap structure would yield a Hebel--Slichter 
peak in all three of these relaxation rates.
We have seen that the transverse relaxation rate $\tau^{-1}$ 
of the $^{63}$Cu nuclei would have 
quite different temperature dependences for the 
$d_{x^2-y^2}$ and $s$--wave gaps.
Magnetic neutron scattering experiments can also provide 
valuable information on the structure of the superconducting gap.
For a wavevector that connects the bonding and antibonding 
fermi surface points,
the scattering intensity at the gap edge is suppressed for an
$s$--wave gap, while we expect to observe a resonance for a 
$d_{x^2-y^2}$ gap structure.

\acknowledgments

The authors gratefully acknowledge support from 
the National Science Foundation under Grant No. DMR95--27304.
The numerical computations reported in this 
paper were performed at the San Diego Supercomputer Center.


\begin{figure}
\centerline{\epsfysize=7.2cm \epsffile[-30 184 544 598] {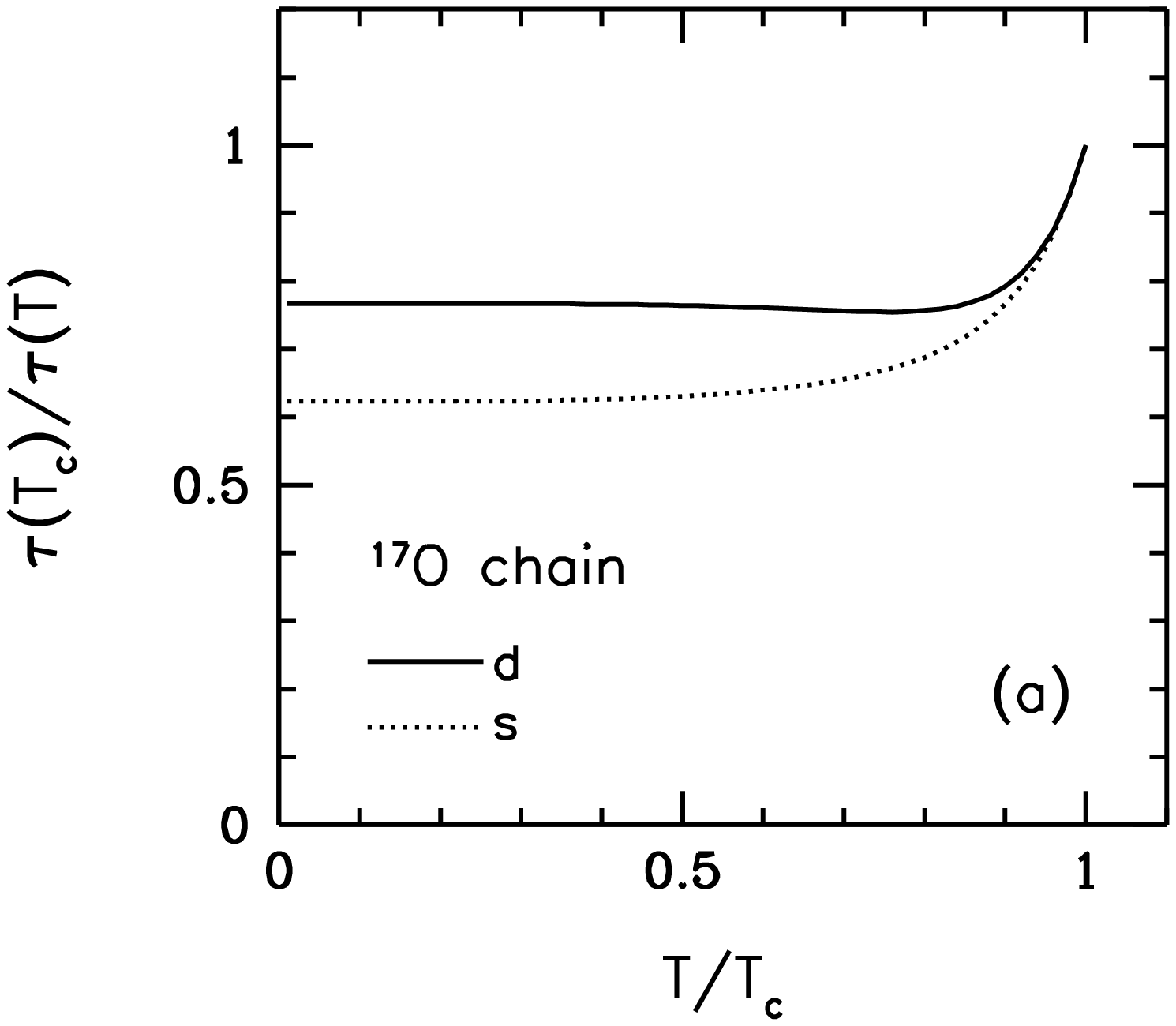}
\epsfysize=7.2cm \epsffile[98 184 672 598] {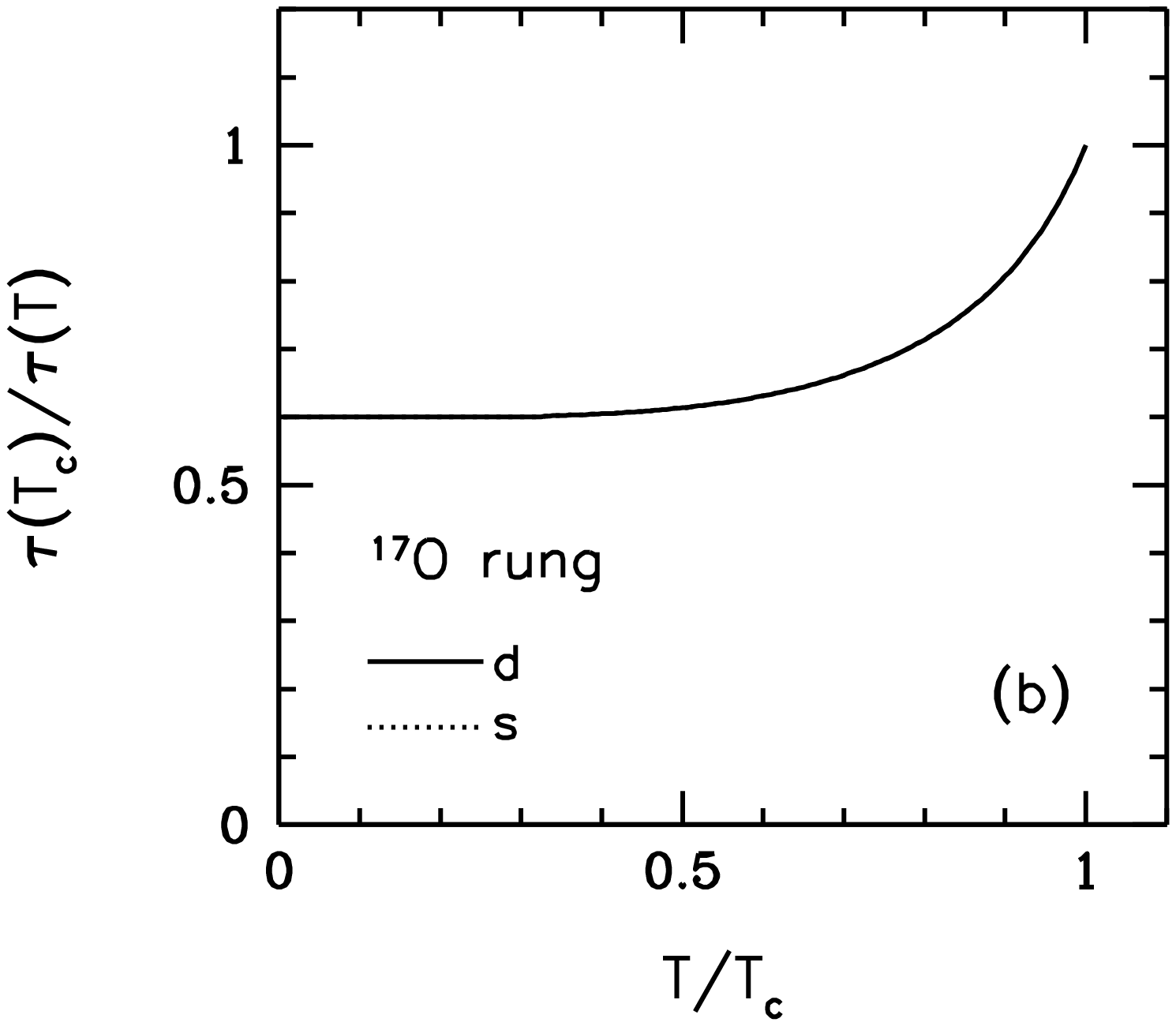}}
\centerline{\epsfysize=7.2cm \epsffile[18 184 592 598] {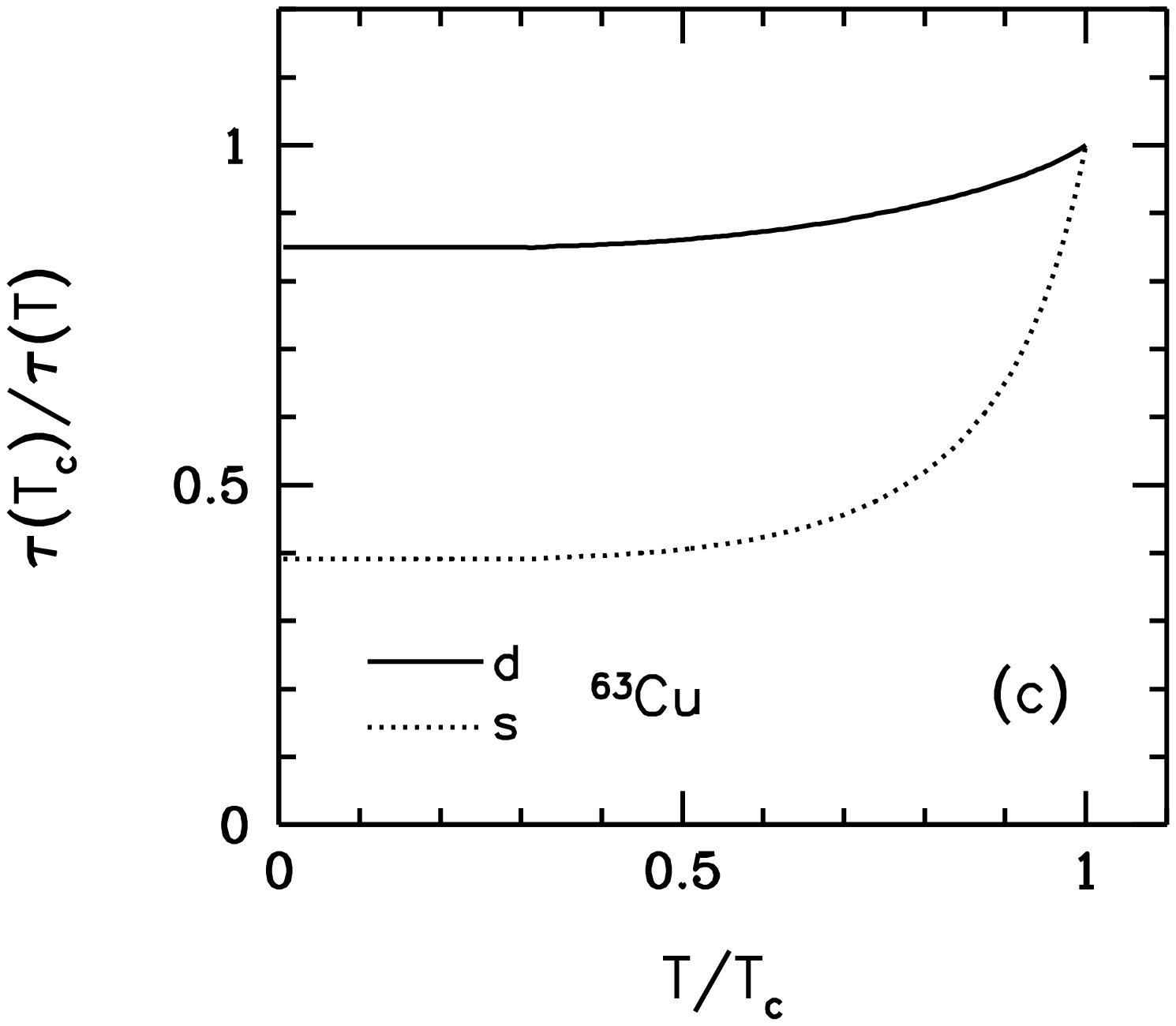}}
\vspace{1.4cm}
\caption{
Temperature dependence of the NMR rate $\tau^{-1}$ for 
(a) the chain $^{17}$O, 
(b) rung $^{17}$O and 
(c) $^{63}$Cu nuclei obtained using the $d_{x^2-y^2}$ and $s$
type of gap structures.
For the rung $^{17}$O nuclei,
the $d_{x^2-y^2}$ and $s$ type of gap structures give 
identical results for the rung $^{17}$O and 
similar results for the chain $^{17}$O.
However, measurements of $\tau^{-1}$ for $^{63}$Cu 
can be used to make a distinction between the two gap structures.
}
\label{fig:rate2}
\end{figure}
  
\begin{figure}
\centerline{\epsfysize=8cm \epsffile[18 184 592 598] {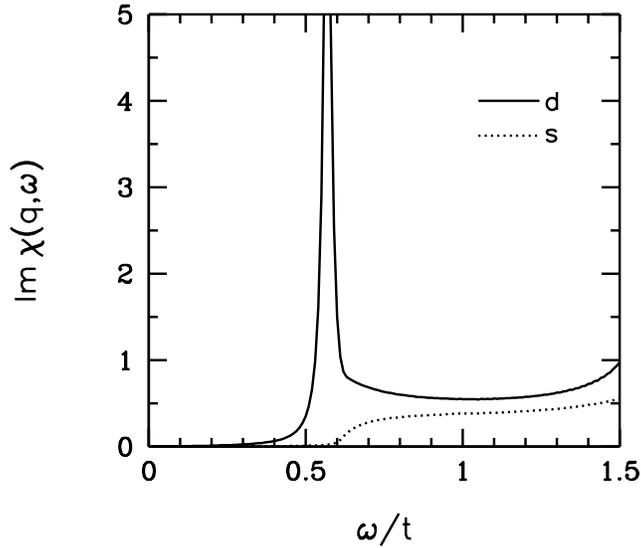}}
\vspace{1.4cm}
\caption{
Inelastic neutron scattering spectral weight 
${\rm Im}\,\chi({\bf q},\omega)$ versus $\omega$ for the
$d_{x^2-y^2}$ and $s$ type of gap structures at $T=0.5T_c$.
Here, ${\bf q}={\bf q}^*=(0.85\pi,\pi)$ is a wavevector 
that connects the bonding and antibonding fermi points.
For an $s$--wave gap, the coherence factor of the quasiparticle 
creation process vanishes and the neutron scattering intensity 
at the gap edge 
$\omega=|\Delta({\bf p}_{{\rm Fa}})|
+|\Delta({\bf p}_{{\rm Fb}})|$
is suppressed.
On the other hand, for the $d_{x^2-y^2}$ gap there is 
a resonance at the gap edge. 
}
\label{fig:imchi}
\end{figure}


\begin{thebibliography}{999}
\bibitem{Like} In the following we will drop the 
qualification ``like'' and use $d_{x^2-y^2}$ to denote 
a gap which has different signs on the bonding and antibonding 
fermi surface points and $s$ to denote a gap which has the same sign.
\bibitem{footnote1} If the value of $\overline{U}$ is larger,
for the $d_{x^2-y^2}$ gap,
the peak in the $T_1^{-1}$ of the chain $^{17}$O nuclei 
can be smaller compared to that of the rung $^{17}$O
because of the faster loss of 
the low--frequency $q_y=\pi$
magnetic fluctuations as the superconducting gap opens.
\bibitem{footnote2} The actual value of the drop in $\tau^{-1}$ 
of $^{63}$Cu depends on the strength of the antiferromagnetic 
fluctuations at $T=T_c$.
For instance, if one uses $\overline{U}=1.8t$ instead of $1.5t$, 
then the system has stronger antiferromagnetic correlations at $T=T_c$
and the drop in $\tau^{-1}$ for the $d_{x^2-y^2}$ gap is 30\%, 
while for the $s$--wave gap it is 75\%.
Measurements of $\tau^{-1}$ and $T_1^{-1}$
for $^{63}$Cu in the normal state will 
determine the strength of the antiferromagnetic fluctuations 
in these materials.
\end{thebibliography}
\end{document}